# Atomic-scale analysis of liquid-gallium embrittlement of aluminum grain boundaries

M. Rajagopalan[1], M.A. Bhatia[1], K.N. Solanki[1*], M.A. Tschopp[2], D. Srolovitz[3]

[1]School for Engineering of Matter, Transport, and Energy, Arizona State University, Tempe, AZ 85281 USA

[2]Dynamic Research Corporation, High Performance Technology Group,

(on site at) U.S. Army Research Laboratory, Aberdeen Proving Ground, MD 21005, USA

[3]Depts. of Materials Science & Engineering and Mechanical Engineering & Applied Mechanics,

University of Pennsylvania, Philadelphia, PA 19104 USA

*+1 (480) 965-1869; +1 (480)727-9321 (fax), E-mail: kiran.solanki@asu.edu, (Corresponding author)

Material strengthening and embrittlement are controlled by intrinsic interactions between defects, such as grain boundaries (GB), and impurity atoms that alter the observed deformation and failure mechanisms in metals. In this work, we explore the role of atomistic-scale energetics on liquid-metal embrittlement of aluminum (Al) due to gallium (Ga). *Ab initio* and molecular mechanics were employed to probe the formation/binding energies of vacancies and segregation energies of Ga for <100>, <110> and <111> symmetric tilt grain boundaries (STGBs) in Al. We found that the GB local arrangements and resulting structural units have a significant influence on the magnitude of vacancy binding energies. For example, the mean vacancy binding energy for <100>, <110>, and <111> STGBs at $1^{st}$ layer was found to be -0.63 eV, -0.26 eV, and -0.60 eV. However, some GBs exhibited vacancy binding energies closer to bulk values, indicating interfaces with zero sink strength, i.e., these GBs may not provide effective pathways for vacancy diffusion. The results from the present work showed that the GB structure and the associated free volume also play significant roles in Ga segregation and the subsequent embrittlement of Al. The Ga mean segregation energy for <100>, <110> and <111> STGBs at $1^{st}$ layer was found to be -0.23 eV, -0.12 eV and -0.24 eV, respectively, suggesting a stronger correlation between the GB structural unit, its free volume, and segregation behavior. Furthermore, as the GB free volume increased, the difference in segregation energies between the $1^{st}$ layer and the $0^{th}$ layer increased. Thus, the GB character and free volume provide an important key to understanding the degree of anisotropy in various systems. The overall characteristic Ga absorption length scale was found to be about ~10, 8, and 12 layers for <100>, <110>, and <111> STGBs, respectively. Also, a few GBs of different tilt axes with relatively high segregation energies (between 0 and -0.1 eV) at the boundary were also found. This finding provides a new atomistic perspective to the GB engineering of materials with smart GB networks to mitigate or control LME and more general embrittlement phenomena in alloys.

# 1. Introduction

Liquid-metal embrittlement (LME) is a phenomenon experienced by many intrinsically ductile metals, including aluminum (Al), nickel (Ni), and copper (Cu). These metals exhibit a drastic loss of ductility in the presence of certain liquid-metals, such as gallium (Ga), bismuth (Bi), and mercury (Hg) [1–10]. Understanding the mechanisms behind LME has been of particular interest in both experimental [2,3,5,8,10–25] and simulation [7,9,26–33] research. Discrepancies between experimental observations and modeling results have led to various proposals for LME mechanisms. One of the most widely accepted LME mechanisms suggests that penetration of the liquid metal into grain boundaries (GBs) modifies interfacial energy so as to promote intergranular failure [3,9,10,20,22–24,29,34,35]. Consequently, the reasons for such variation is likely associated with the specific type of liquid environments [10,18,29] tested, whereby some elements act to enhance cohesion in certain systems, while others promote decohesion [28,35]. However, recent nanoscale experimental studies suggest that one of the primary contributing factors in LME is the formation of intermetallic compounds at GBs [1] and bilayer interfacial phases (e.g., in Ni-Bi) [17]. The lack of a definitive understanding of the mechanistic origin of LME hinders our ability to satisfactorily address LME in important technologies, including in the nuclear-power generation sector, where liquid-metal coolants are in contact with metallic structural components, and in industrial sectors that employ liquid-based joining technologies (e.g., soldering and brazing) [4,9,10,16,19,32].

Al-Ga is often viewed as a model LME system. Al and Ga are from the same chemical group (IIIA), have similar electronegativities, and are of comparable atomic size [36]. Hence, they are expected to have similar chemical properties [27]. Nonetheless, liquid Ga penetration along Al grain boundaries is rapid and results in a substantial loss of cohesion [32]. The effectiveness of LME in this system is known to depend on grain size [37]. Several experimental techniques have been employed to reveal the atomic-scale mechanisms that give rise to LME in this system. High resolution transmission electron microscope (TEM) has provided important insight into LME in Al-Ga [22–24]. While Ga does penetrate most GBs in Al, leaving behind Ga layers that range from one to several monolayers [9,22–24,27], *in situ* TEM studies have shown that low energy boundaries, such as high coincident site density (small $\Sigma$) boundaries, show little segregation [23]. The interplay of the variables that influence embrittlement complicates the separation of which properties affect and do not affect LME. The difficulties with preparing and fully characterizing experimental samples (especially their GB bicrystallography), has made the prospect of accurate simulations appealing. Atomic-scale simulations offer advantages for the analysis of grain boundary properties, albeit at very small length and time scales. Nonetheless, such simulations can help establish the relationship between embrittlement and material/GB properties.

Atomistic and quantum mechanics-based simulations are increasingly utilized in investigations of fundamental LME mechanisms. Stumpf and Feibelman [27] employed first principle methods to study Al-Ga and they showed that Ga lowers the surface energy of Al. This is important since this suggests that Ga may lower the cohesive energy of interfaces in Al. The attraction of Ga to substitutional sites in Al GBs was demonstrated by Thomson *et al.* [26] using *ab initio* calculations for $\Sigma 5(310)$ symmetric tilt grain boundaries (STGBs). They concluded that local strain governs the substitution characteristics. These efforts and others elucidate that density functional theory (DFT) is effective for characterizing the electronic origins of LME. However, the computational expense associated with such calculations has limited their application to special GBs with a small number of atoms per cell. Monte Carlo and molecular dynamics (MD) methods have also been widely used to characterize LME [2,5,9,16,23,25,38,39]. While such calculations commonly incorporate a much larger numbers of atoms, they have been limited by their use of empirical descriptions of atomic interactions. Such potentials are typically fit to *ab initio* data and relevant experiments. Most of the early atomistic work quantified the role of GB structure on Ga penetration both with and without external stresses, suggesting that in some bicrystal systems, Ga penetration did not occur without application of external stimuli [5,23]. These studies indicate that GB characteristics play a significant role in embrittlement [39]; leading to considerable interest in correlating GB structure with macroscale embrittlement. For example, MD simulations suggest that the atomistic mechanisms of LME are sensitive to boundary structure [29]. In another example, Nam and Srolovitz [32] found that Ga penetration in Al was smaller along symmetric GBs compared with asymmetric GBs and that special low-$\Sigma$ GBs are, indeed, special. The sensitivity of LME to GB structure and the fact that most studies focused on special boundaries provides impetus to explore how variations in structure and properties among large sets of general GBs influence LME.

Grain boundaries are two-dimensional defects which are characterized by five macroscopic degrees of freedom [40–48]. Saylor *et al.* [40] analyzed the GB character distribution (GBCD) in commercially pure Al and suggested that boundaries with low index planes occur with particularly high frequency in polycrystals. The GB structure-energy correlation in several fcc metals was investigated by Wolf [49] who established a linear relationship between GB energy and volume expansion per unit area (grain boundary free volume). The role of the GB plane in determining GB energy was investigated by Holm *et al.* [43], who performed extensive calculations of GB energies in Ni and Al. Rohrer [50] established that the coincidence lattice site density ($\Sigma$) plays only a minor role as a determinant of GB energy in several systems and that the energies of the crystal surfaces that meet at a GB is a better indication of GB energy. Tschopp and McDowell [51] showed that asymmetric tilt boundaries in Cu and Al, often decompose into lower energy, symmetric tilt orientations via faceting. Hasson *et al.* [52] experimentally measured GB energies for a wide range of <100> and <110> symmetric tilt boundaries in Al and concluded that lower energy boundaries show unique segregation, corrosion, and diffusion behavior.

In this paper, we perform a systematic investigation of LME in the Al-Ga system using *ab initio* pseudopotential calculations and molecular statics (MS) simulations to clarify the relation between GB structure, Ga segregation, and GB cohesive energy – a quantity with LME fracture implications. Here, the GB database consists of various <100>, <110>, and <111> symmetric tilt GBs (125 GBs). The paper is organized as follows. The next section describes the computational methods and GB generation procedures employed. Then, we present several different characterizations of GBs with implications for LME: (1) metrics that correlate local GB structure with atomic-scale segregation and embrittlement phenomena (e.g., GB energy and free volume), (2) vacancy binding energy at the GB, and (3) the energies associated with Ga segregation to Al grain boundaries. We show that GB structure plays a significant role in Ga segregation and the subsequent embrittlement of Al. Examination of vacancy binding energies as a function of distance from the GB plane shows that GBs do not always act as vacancy sinks. Ga segregation energies are strongly anisotropic and this anisotropy is strongly correlated with GB structure. The thickness of the Ga segregated layer in the GB varies from 4-6 atomic planes, depending on tilt axis. This, together with the Ga diffusion distances, provide necessary length-scale parameters for potential inclusion in larger-scale models, such as phase-field simulations [53].

Finally, an atomistic simulation framework is developed that addresses the stochastic nature of GB segregation and embrittlement in LME environments. This framework provides guidance for engineering the GB character distribution in polycrystals.

## 2. Computational methods

To investigate GB sink efficiency and subsequent embrittlement in Al-Ga, we employed molecular static (MS) simulations using an embedded atom method potential and *ab initio* calculations based upon a pseudopotential method within a generalized gradient approximation (GGA). 125 STGBs were created in Al (<100>, <110> and <111> tilt axes) and equilibrated using MS simulations performed with the Large-scale Atomic/Molecular Massively Parallel Simulator (LAMMPS) [54]. The simulation cell consisted of two single crystal layers meeting at a planar grain boundary with periodic boundary conditions in all three orthogonal directions [45,51,55–61]. The thickness of the simulation cell in the direction perpendicular to the boundary plane was ~12 nm (this large value was chosen in order to eliminate any effects associated with the periodic boundary condition). Several 0 K minimum-energy GB structures were obtained through successive rigid body in-plane translations of the two grains followed by application of an atom-deletion technique and energy minimization using a conjugate gradient method [45,51,54,55,57,58,60–64]. The embedded atom potential (EAM) developed by Nam and Srolovitz [9] was used to describe the Al-Al interactions in generating the impurity-free GB systems and to describe Al-Ga interaction in the segregation calculations. This potential was parameterized using a database of energies and configurations (from DFT calculations) including heats of solution and the diffusivity of Ga in bulk Al, the binding energy of Ga to free surfaces, vacancies, and dislocations, as well as other quantities that depend on the Ga-Al interaction in order to describe the properties of the binary Al-Ga system. These potentials yield good agreement with the experimental measurements of the self-diffusivity and Al diffusivity in liquid Ga calculated. These STGBs were used as input to perform further MS and DFT calculations. *Ab initio* calculations were performed on a representative set of GBs generated using MS (at 0 K) as input to the vacancy and segregation calculations. The bicrystal simulation cells employed in the DFT calculations were performed on smaller unit cell than in the static relaxation calculations (these too were first relaxed using the same EAM potential).

DFT calculations were performed using the Vienna Ab-initio Simulation Package (VASP) plane wave electronic structure code [65–67]. Projector augmented wave (PAW) [68,69] potentials were used to represent the nuclei and core electrons up to the 2p shell for Al. Exchange and correlation was treated with GGA using the PBE [70] form with an energy cutoff of 312.39 eV and the Monkhorst Pack k-point mesh given in Table I. The GB structure obtained using the EAM potential was further relaxed using the VASP code with a conjugate gradient algorithm [71] with 1 meV/Å force and 0.01 meV energy convergence criteria. A single vacancy was introduced at different distances from the grain boundary and the vacancy formation energy was calculated following energy minimization. This process was repeated for single vacancies in different positions parallel to the boundary plane.

TABLE I. Characterization of aluminum STGBs from MS and DFT simulations.

| GB Σ (plane) <axis> | K-Points | X x Y x Z (Å) | # Atoms in DFT Cell | GB Energy (mJ/m$^2$) MS | GB Energy (mJ/m$^2$) DFT | GB Free Volume/Area (Å) |
|---|---|---|---|---|---|---|
| 5 (210) <100> | 11 x 7 x 15 | 9.04 x 28.05 x 4.04 | 60 | 565 | 555 | 0.0945$a_0$ |
| 5 (310) <100> | 11 x 7 x 15 | 12.78 x 25.21 x 4.04 | 76 | 551 | 553 | 0.1670$a_0$ |
| 13 (320) <100> | 9 x 7 x 15 | 14.57 x 28.13 x 4.04 | 98 | 481 | 480 | 0.225$a_0$ |
| 13 (510) <100> | 11 x 5 x 15 | 10.30 x 40.39 x 4.04 | 100 | 542 | 548 | 0.1754$a_0$ |
| 3 (111) <110> | 11 x 7 x 13 | 9.90 x 25.77 x 5.71 | 88 | 11 | 13 | 0 |

The effect of GB character on vacancy sink efficiency was assessed by calculating the formation/binding energies for vacancies in two situations: (1) using both MS and *ab initio* calculations in which a vacancy was placed in atomic layers at different distances from the GB plane (the 0$^{th}$ layer represents the GB plane) and (2) using MS calculations in which a vacancy was placed at different sites within ±15 Å of the GB plane. The calculations helped identify the variation of vacancy formation/binding energies within each layer and the effect of GB structure on vacancy formation energies in a few, selected GBs. Altogether, this entailed ~150,000 MS calculations to quantify the vacancy formation/binding energy statistics. The vacancy formation energy at a site α is $E_f^\alpha = E_{Gb}^\alpha - E_{Gb} + E_{coh}$, where $E_{coh}$ is the cohesive energy/atom in a perfect fcc lattice, and $E_{Gb}^\alpha$ and $E_{Gb}$ are the total energies of the GB simulation cell with and without the vacancy, respectively [72,73]. It is useful to reference the vacancy formation energy at the grain boundary $E_f^\alpha$ to that in the bulk $E_f^0$

$$E_b^\alpha = E_f^\alpha - E_f^0 \tag{1}$$

in order to assess the energy required to move the vacancy from the boundary into the bulk or vice versa. The bulk vacancy formation energy $E_f^0$ is ~0.88 eV, as determined from the MS calculations and ~0.72 eV from the *ab initio* calculations.

The effect of GB structure on segregation and embrittlement behavior was examined. The first step was to determine the energy required to substitute a Ga for Al atom in bulk Al. The energy to segregate a Ga to a GB in Al was calculated using the same two approaches used to determine vacancy segregation (as described above). The energy was minimized after replacing one Al atom with a Ga atom following which the segregation energies were calculated as

$$E_{Seg}^\alpha = (E_{GB}^\alpha - E_{GB}) - (E_B^0 - E_B) \tag{2}$$

where $E_{GB}^\alpha$ and $E_{GB}$ are the total energies of the GB simulation cell with and without a Ga atom substituted for an Al atom at site α, respectively. Similarly, $E_B^0$ - $E_B$ is the difference in energy between bulk Al in which one Ga replaces one Al atom and the same single crystal of all Al atoms. This expression for the segregation energy can be thought of as the energy required to move a single Ga atom from a position far from the grain boundary to position α in the vicinity of the grain boundary [57,60,74]. Negative values of the segregation energy implies that segregation is energetically favorable, while positive values indicate that it is favorable for Ga to leave the boundary and go into solution.

The work of interfacial adhesion or cohesion, $2\gamma_{int}$, can be thought of in terms of ideal fracture. It is the change in energy of a system with an interface if a perfectly planar crack propagated along an interface cleaving in two assuming that there is no energy dissipation. In other words it is the difference in energy between a system with a grain boundary and the system if it were to split apart a grain boundary to form two free surfaces, i.e.,

$$2\gamma_{int} = 2\gamma - \gamma_{gb} \tag{3}$$

where $2\gamma$ is the energy (per area) of the two free surfaces and $\gamma_{gb}$ is the energy of the grain boundary. If Ga atoms were substituted for Al atoms at sites α in the vicinity of a grain boundary, the work of adhesion or cohesive energy of the boundary would be

$$2\gamma_{int}^{\alpha} = 2\gamma^{\alpha} - \gamma_{gb}^{\alpha} \qquad (4)$$

where $2\gamma_{int}^{\alpha}$ is the work of adhesion of the interface with Ga atom at sites α, $2\gamma^{\alpha}$ is the energy of the two free surfaces after separation with Ga in sites α and $\gamma_{gb}^{\alpha}$ is the GB energy with Ga at α.

The embrittling effect of Ga segregation to a grain boundary can be judged by comparing the work of interface adhesion with and without Ga segregated to the grain boundary. In other words if $2\gamma_{int}^{\alpha} < 2\gamma_{int}$, it is easier to fracture a Ga segregated boundary than the same boundary in the absence of Ga segregation. This can also be assessed by directly inspecting the segregation energies of the free surface and the GB. A difference in segregation energies indicates the ability of the atom to reduce or enhance GB cohesion [75,76].

## 3. Grain boundary structure and energy

Understanding the structure and energy of GBs is crucial for engineering materials resistant to LME since the embrittlement potency is boundary (and environment) specific. We report here on GB structure and energy, vacancy formation energies, and Ga substitution energies in Al for a wide range GBs; i.e., 125 STGBs with <100>, <110> and <111> tilt axes in Al. The GB energies as a function of the misorientation angle for the <100> STGB configurations are compared to previously reported GB energies to assess the Al portion of the Al-Ga interatomic potential, see Fig. 1(a). The GB energy versus misorientation angle trend is very similar to those of Tschopp and McDowell [51] (Mishin *et al.* [77] EAM potential) and Hasson *et al.* [52] (Morse potential). We find that changing the interatomic potential results only in small changes in the GB energies for the symmetric tilt systems.

The misorientation-energies were mapped onto a stereographic triangle appropriate for cubic metals (see Fig. 1(b) [49]) where the corners of the triangle represent the three principal orientations of the cubic system and the angle associated with each GB for the <100>, <110>, and <111> tilt systems was expressed by polar/azimuthal angle pairs. In Fig. 1(b), we interpolate the GB energy data using a contour plot, where the colors represent grain boundary energy.

Representation of the entire data set on one plot serves as an aid for identifying targets for GB engineering.

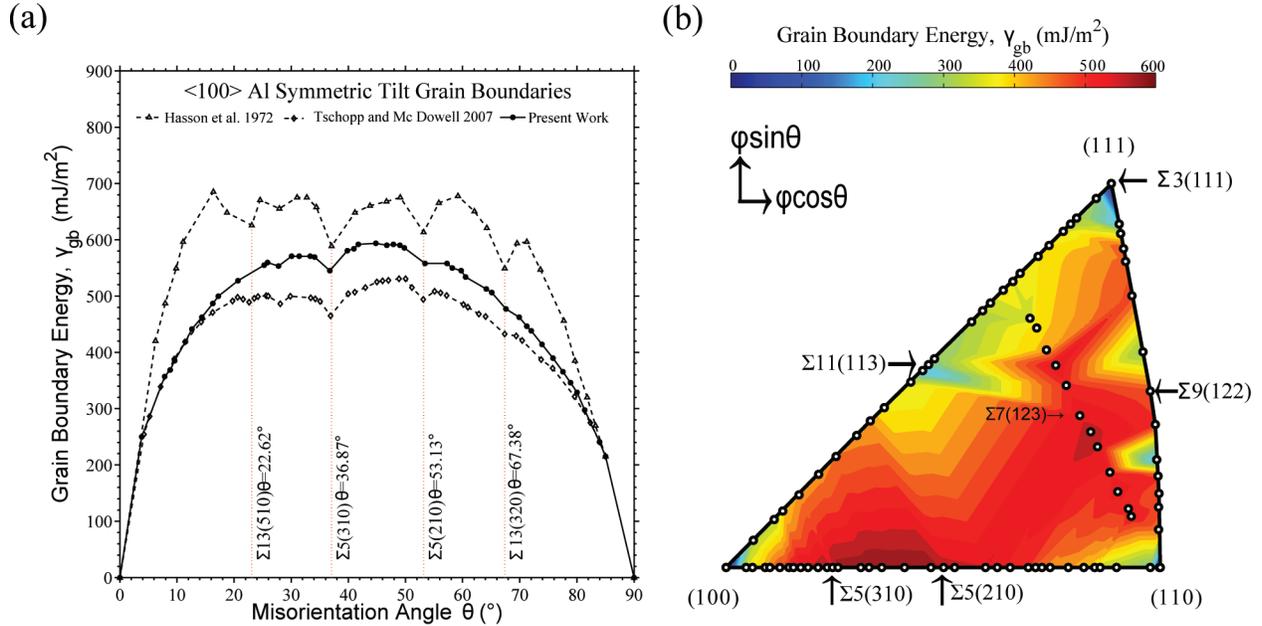

Fig. 1. (a) Plot of GB energy versus GB misorientation angle for Al STGBs with a <100> tilt axis. The energy cusps are identified with dotted red lines. The data presented are from the present work as well as Tschopp and McDowell [51] and Hasson *et al*. [52] (b) Contour plot of GB energies for the three symmetric tilt systems (<100>, <110>, and <111>) in Al, represented using polar and azimuthal angles. This plot is similar to a stereographic projection of the planes in a cubic crystal. The polar and azimuthal angles are degrees of freedom characterizing the GB plane (hkl). The energy cusps for each tilt system are labeled.

The structure-energy correlation can be analyzed with the aid of structural unit characterizations of the local atomic configurations in the GB. For low-angle boundaries, these units describe discrete dislocations. However, at higher misorientation angles (high-angle GBs), the dislocation cores overlap, dislocations relax to minimize the boundary energy, yet such GBs can be characterized by GB dislocations or structural units [45]. In certain GBs (typically low Σ), only a single "favored" structural unit appears, while other boundaries are characterized by structural units from the two neighboring favored boundaries. The GB structures in the <100> STGB system are shown in Fig. 2. For instance, the structural units in the Σ5(310)θ=36.87° and Σ5(210)θ=53.13° STGBs are labeled as B' and C, respectively, and Σ13(510)θ=22.62° GB is represented as |CDD|. The Σ3(112)θ=70.53° GB, the incoherent twin boundary, is a combination

of the C and D structural units from the Σ11(113)θ=50.48° GB (C) and the Σ3(111)θ=109.47° (D) coherent twin boundary, respectively. The structural units reported here are represented in a convention widely used to represent fcc cubic metals [45,58,59,78]. The Voronoi atomic volume $V_{vor}$ is also computed for the boundaries in Fig. 2; the Voronoi volume is largest at the GB center and converges to the bulk Voronoi volume (16.6 Å$^3$) as distance from the GB increases. Interestingly, the <100> STGBs exhibit higher Voronoi volumes that the <110> STGBs, due to the larger interplanar spacing in the <100> tilt direction. This GB metric can in turn be correlated with other energetics associated with the GBs to derive structure-property relationships.

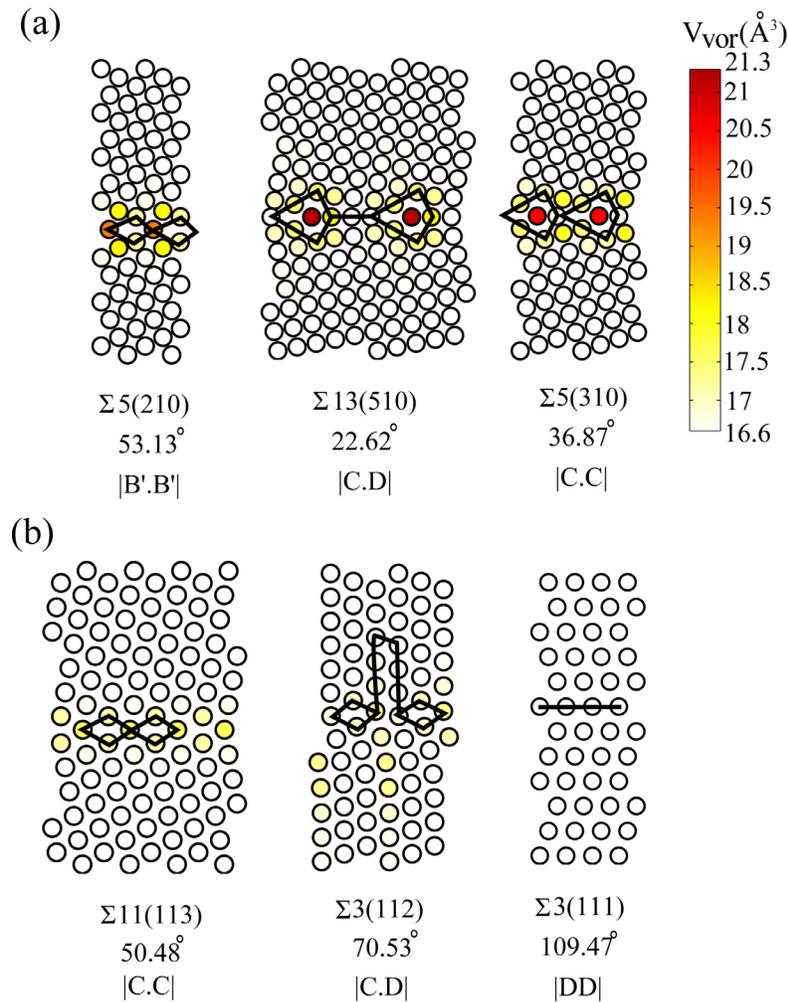

Fig. 2. Structural units and Voronoi atomic volumes for (a) <100> STGBs: Σ5(310)θ=36.87°, Σ13(510)θ=22.62°, and Σ5(210)θ=53.13° GBs. The structure of Σ13(510)θ=22.62° GB is a combination of C and D units. The structure and Voronoi volumes for the <110> STGBs: Σ11(113)θ=50.48°,

Σ3(112)θ=70.53°, and Σ3(111)θ=109.47° GBs are shown in (b). The structure of Σ3(112)θ=70.53° is a combination of the units C and D.

## 4. Vacancy binding

Vacancies play an important role in segregation and embrittlement through substitutional solute kinetics and transport in and along the grain boundary. Therefore, before turning to Ga segregation *per se,* we first focus on vacancies near the grain boundary. Molecular statics and *ab initio* calculations were used to examine the vacancy binding energy as a function of local atomic structure, distance from the GB plane, and at different layers parallel to the GB plane. Molecular statics simulations were used to quantify the statistical nature of the vacancy formation energy for all sites within 15 Å of the GB center for all the GB systems (~150,000 MS simulations). The change in the vacancy binding energies with distance from the GB center can be used to quantify the length scale of the vacancy-GB interaction. Figure 3(a) shows results for the case of <100> STGBs in Al; we observe similar behavior for <110> and <111> STGBs. The majority of the GB sites have vacancy binding energies that are more negative than the bulk value and the sites away from the GB have vacancy formation energies similar to the bulk value.

To quantify the spatial extend of the binding energy relative to the bulk, we focus on the full width half maximum (FWHM); i.e., the width of the distribution at which the data falls to half its peak value. This indicates the length scale over which the probability of finding sites with large negative vacancy binding energies is high (~68%). This technique was adopted to rigorously identify the vacancy segregation length scale for <100>, <110> and <111> STGBs. In the case of <100> GBs, the FWHM is approximately 10 Å, or within approximately 5Å of the GB plane. Further, we see that this vacancy/GB interaction length is smaller for high angle than low angle boundaries, which is not surprising since the strain fields associated with the GB are longer range for small misorientations than for large misorientations. Figure 3(b) illustrates the binding energy variation for a high and a low angle boundary. The same methodology was applied to calculate the FWHM for the <110> and <111> STGBs and the corresponding length scales were identified as 14 Å and 12 Å, respectively. In other words, the length over which there is significant grain boundary/vacancy interaction is 5 Å, 6 Å and 7 Å from the GB plane for <100>, <111> and <110> tilt axis boundaries, respectively. In order to identify the distribution of vacancy binding energies within the GB region, a few <100> STGBs were chosen for further analysis. Figure 3(c) shows the spatial distribution of the vacancy binding energies for selected GBs. These figures show that the local environment strongly influences vacancy formation/binding energies.

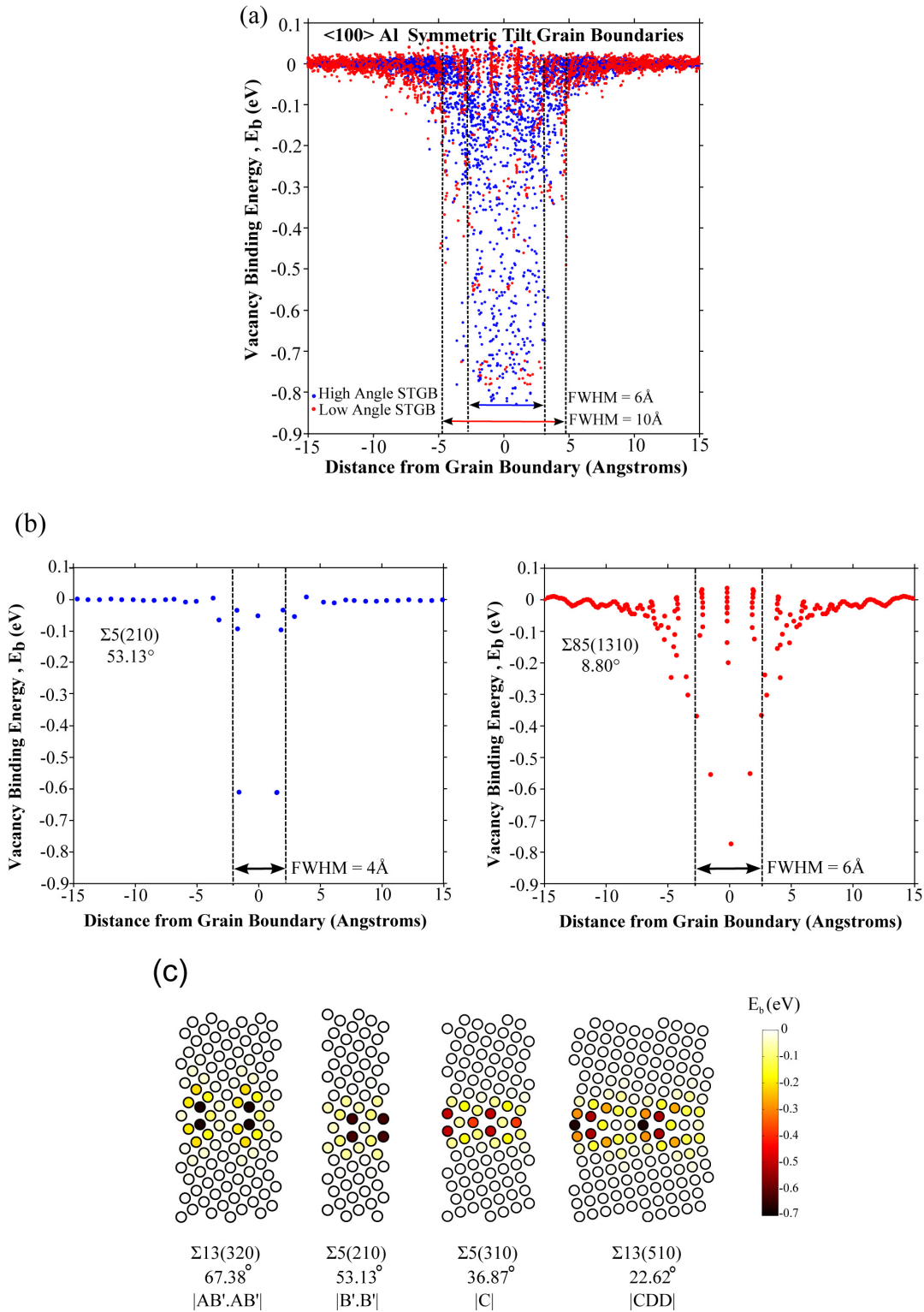

Fig. 3. Vacancy binding energy $E_b$ as a function of distance from the GB plane for (a) a wide range of <100> STGBs and (b) a single high angle boundary $\Sigma 5(210)<100>$ and a single low angle boundary $\Sigma 85(13,1,0)<100>$, and (c) the site-specific vacancy binding energy in the vicinity of several <100> STGBs. The strongest binding is for sites close to the GB plane for both high and low angle GBs, although this effect is stronger for high angle GBs than low angle GBs. The interaction length scale for the low angle boundaries is higher than that of the high angle boundaries.

In order to obtain a more averaged view of how the vacancy binding energy varies with distance from the GBs, we report the vacancy binding energy data distribution for all 50 <100>, 50 <110> and 25 <111> STGBs in Fig. 4; we plot the binding energy versus distance from the grain boundary, normalized by the separation between planes parallel to the GB, i.e., $d/d_{\{hkl\}}$. This effectively counts the number of atomic planes from the GB center. Figure 4 shows a boxplot representation of the vacancy binding energy distribution on each plane. The binding energy of the vacancy to sites near the GB center (1$^{st}$ atomic plane) for the <100> and <111> systems (-0.63 eV and -0.60 eV, respectively) is lower than that at the <110> STGBs (-0.26 eV), implying that overall <100> and <111> STGBs are much more effective vacancy sinks than <110> STGB. However, all of the STGBs studies are preferred vacancy sites as compared with the bulk. In all cases, the vacancy binding energy approaches zero after ~6 atomic planes (although the decay is somewhat slower for <110> STGBs).

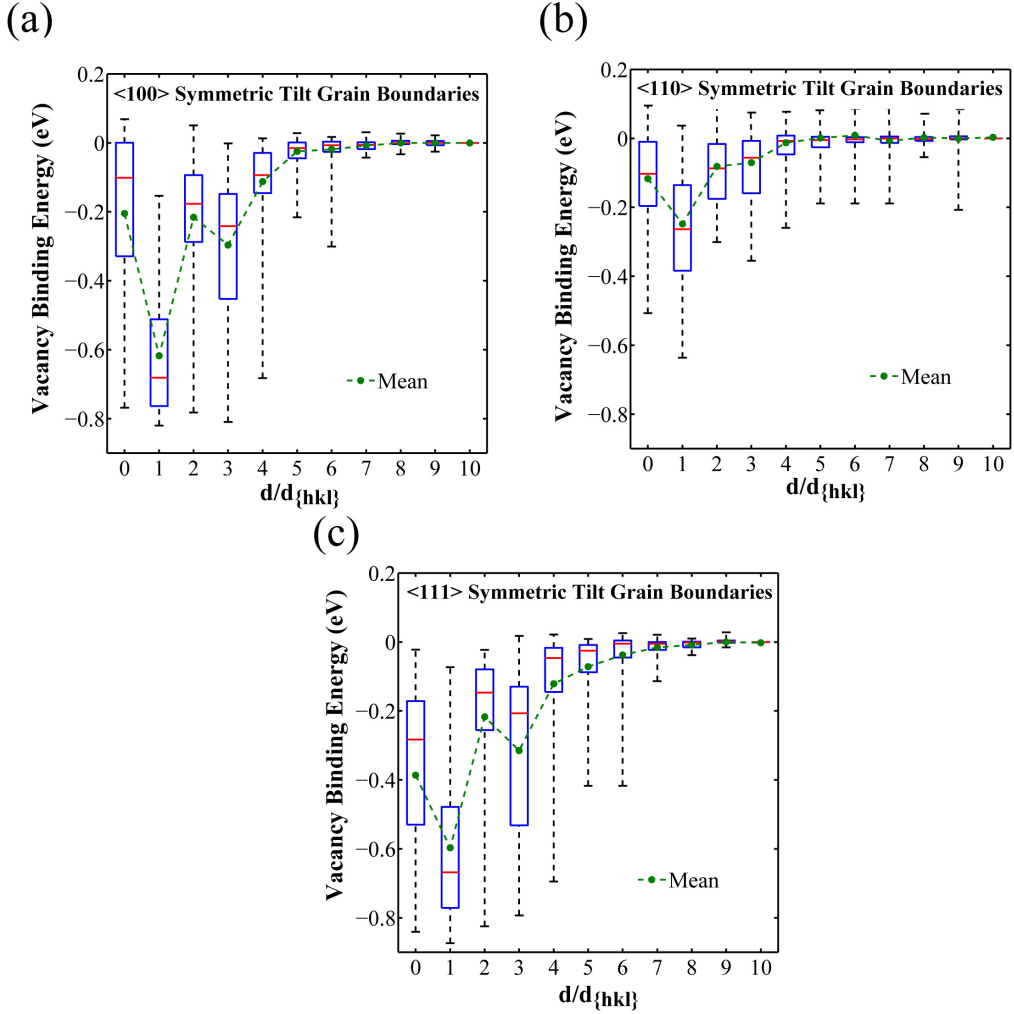

Fig. 4. The vacancy binding energies $E_b$ versus the normalized distance from the GB center $d/d_{\{hkl\}}$ for <100>, <110> and <111> Al STGBs. This is equivalent to counting the number of atomic planes (parallel to the GB plane). The plot shows the mean (green dot) and median (red bar) of the vacancy binding energy distribution, as well as 25th and 75th percentiles (bottom and top of the blue boxes) and the extreme minimum and maximum energies (ends of the black dotted lines).

We now compare the EAM MS vacancy binding energy results with those obtained using DFT (relaxed configurations). We note that while the vacancy formation energy in the perfect crystal is $E_f = 0.88$ eV (EAM MS) and $E_f = 0.72$ eV (DFT). The results are shown in Fig. 5 for two $\Sigma 13$ and two $\Sigma 5$ <100> STGB and the twin $\Sigma 3$ <110> STGB. While the trends in the vacancy binding versus distance data are in excellent agreement between the two methods, there is a slight tendency for the DFT results to be somewhat higher than those from the EAM MS calculations. This is either due to the differences in the description of binding or to the fact that the DFT calculations were performed on smaller simulation cells (more constraint). The data also show while there is a tendency for vacancies to strongly bind to <100> STGBs, this is strongest one atomic plane away from the STGB center. We note that binding energy of a vacancy to the coherent twin boundary Σ3(111)<110> is essentially zero at all separations. This is a result of the fact that the atomic configuration at the twin boundary is nearly identical with that of the perfect crystal from the point of view of free volume and local packing and is not representative of more general GBs. One interesting observation is that the vacancy binding energy anisotropy (i.e., the difference in energy between the vacancy on the $0^{th}$ layer and the $1^{st}$ layer of a GB) is inversely related to the GB free volume. For example, the Σ5(210) θ=53.13° GB has an initial free volume of $0.0945a_0$ and a high degree of anisotropy (see Fig. 5) when compared to Σ5(310)θ=36.87° GB, which has an initial free volume of $0.167a_0$ and negligible vacancy binding energy anisotropy (in Al, $a_0$ = 4.05 Å) [9].

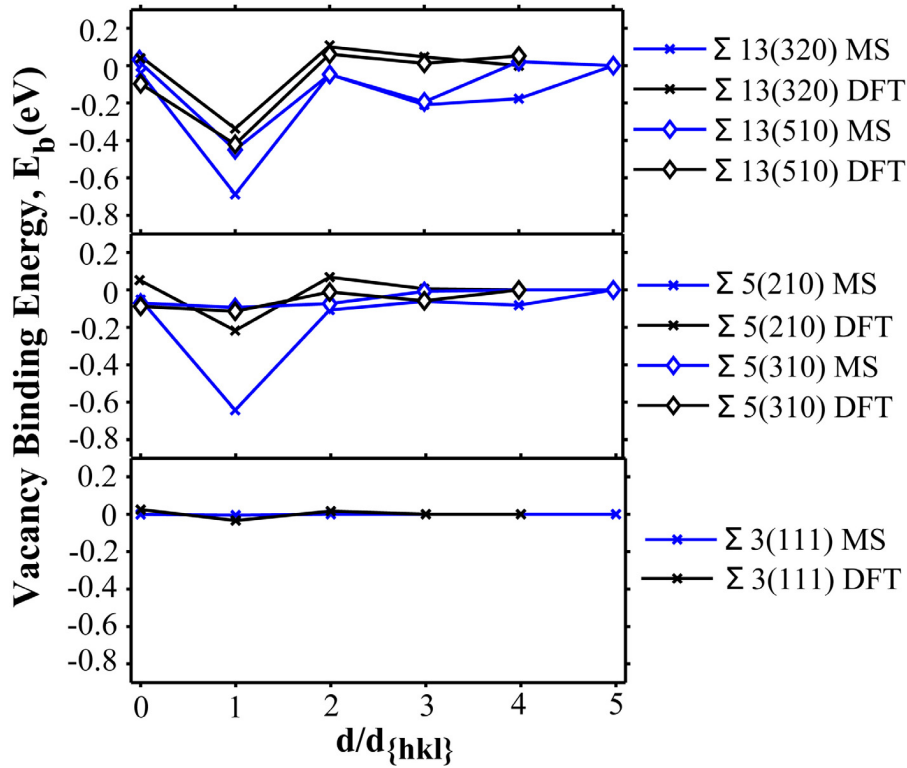

Fig. 5. Vacancy binding energy as a function of normalized distance from the GB for selected <100> STGBs and the coherent twin Σ3(111)<110> Al STGBs calculated using EAM MS and DFT simulations. The dotted lines represent length scales from GB center plane.

## 5. Ga segregation

Molecular statics and *ab initio* techniques are employed to examine the segregation of Ga to GBs in Al at sites near the GB as well as a function of distance from the GB plane. These calculations are designed to identify why some GBs in Al are wetted by Ga and other not and why in some cases the wetting layer is only a monolayer while in others it is multiple layer thick [17]. As with vacancies, MS calculations were used to quantify the statistical aspect of segregation for all atomic sites within 15 Å of the GB for all 125 STGB (~150,000 simulations). Several GBs were then selected for closer examination based upon DFT calculations.

Table II illustrates the variation of GB segregation energies for several low Σ STGBs as a function of normalized distance $d/d_{\{hkl\}}$ (i.e., plane number) computed using MS and DFT for a few selected GBs which correspond to local minima in the misorientation-energy plot of <100> and <110> symmetric tilts, see Fig. 1(b). While Ga tends to segregate to the GB plane ($d/d_{\{hkl\}}=0$) (negative segregation energies) this tendency is even stronger at the next atomic plane $d/d_{\{hkl\}}=1$ (larger magnitude, negative segregation energies). While this general trend holds for most of the GBs, it fails for the Σ3(111) GB which shows almost zero Ga segregation energy throughout the bicrystal. There appears to be a correlation between the variation of the Ga segregation energy with lattice plane and that for the GB free volume; compare Tables I and II. For example, while the difference in segregation energies between the GB plane and the next plane for the Σ5(210) GB is 0.168 eV (EAM MS) with a free volume of $0.0945a_0$, the Σ3(111) GB exhibits almost zero free volume and the difference in the Ga segregation energies between the same two planes is 0.01 eV. We note that there is a small difference in the segregation energies calculated using DFT and EAM, however the errors are not large and the variation in segregation energy with plane tracks the same way in both.

The Ga preference for the GB or the bulk was further quantified by computing the segregation energy as a function of distance from the grain boundary for 125 STGBs in three tilt systems. Figure 6 plots the segregation energy as a function of normalized interplanar distance for the <100>, <110>, and <111> STGBs. In all cases, Ga tends to segregate to the GBs and this tendency is stronger one atomic plane from the GB plane than to the GB plane itself. These results indicate a strong correlation between Ga segregation and the local Al GB structure and could lead to GB decohesion as reported in Ludwig and Bellet [16] and in Nam and Srolovitz [32]. There also is a small, second minima in the segregation energies at $d/d_{\{hkl\}}= 3$ in the <100> and <111> tilt axis GBs, the origin of which is not understood.

TABLE II. Gallium segregation energies for selected GBs obtained using EAM MS and DFT calculations.

| GB Σ (plane) <axis> | $d/d_{\{hkl\}}=0$ Seg. Energy (eV) | | $d/d_{\{hkl\}}=1$ Seg. Energy (eV) | | $d/d_{\{hkl\}}=2$ Seg. Energy (eV) | | $d/d_{\{hkl\}}=3$ Seg. Energy (eV) | |
|---|---|---|---|---|---|---|---|---|
| | MS | DFT | MS | DFT | MS | DFT | MS | DFT |
| Σ5 (210) <100> | -0.07 | -0.14 | -0.24 | -0.16 | -0.07 | -0.13 | 0.00 | 0.00 |
| Σ5 (310) <100> | -0.17 | -0.11 | -0.22 | -0.17 | -0.19 | -0.02 | 0.02 | -0.01 |
| Σ13 (320) <100> | -0.07 | -0.09 | -0.22 | -0.10 | -0.03 | 0.01 | -0.12 | -0.11 |
| Σ13 (510) <100> | -0.09 | -0.09 | -0.14 | -0.16 | -0.06 | -0.12 | -0.01 | 0.00 |
| Σ3 (111) <110> | 0.00 | -0.02 | 0.01 | 0.00 | 0.00 | 0.00 | 0.00 | 0.00 |

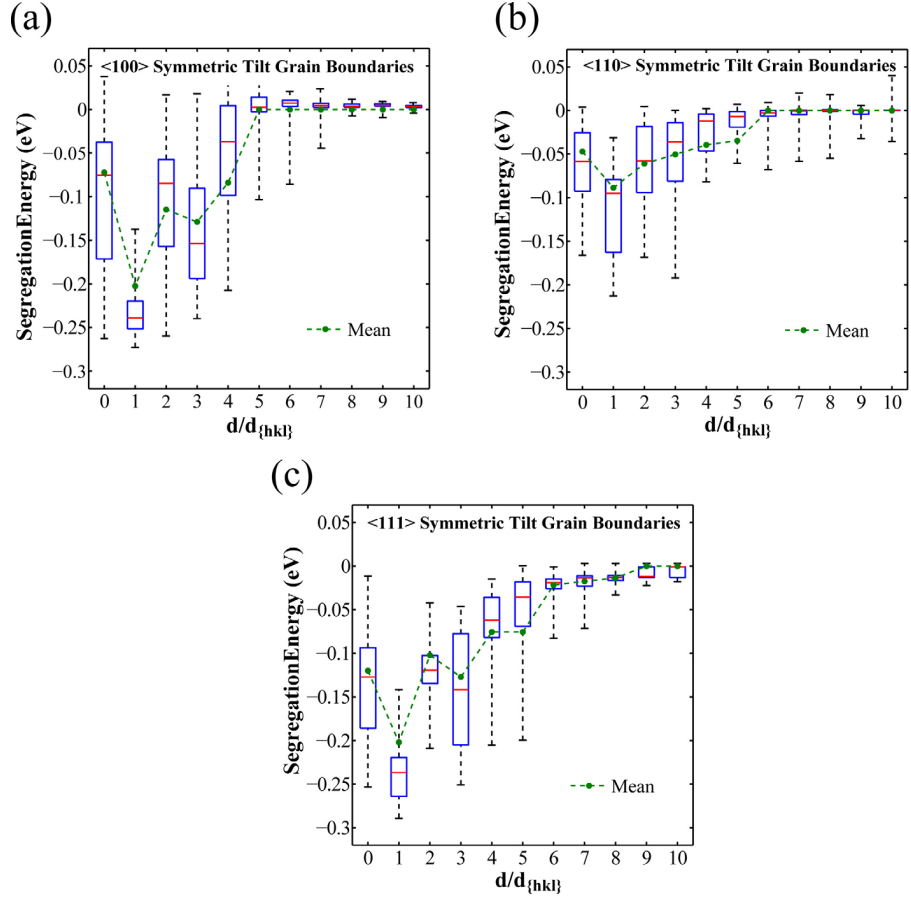

Fig. 6. Segregation energy as a function of distance $d/d_{\{hkl\}}$ from the GB plane for <100>, <110>, and <111> tilt axes. These plots are in the same format as Fig. 4.

The site-specific Ga segregation energies for four STGBs in Al are illustrated in Fig. 7; the associated structural units and Voronoi volumes were shown in Fig. 2. While there is a tendency for Ga segregation to all of the STGBs, this tendency is particularly weak for the $\Sigma 3(111)<110>$ and $\Sigma 3(112)<110>$ GBs. On the other hand, $\Sigma 5(210)<100>$ and $\Sigma 5(310)<100>$ GBs have a strong segregation tendency ($E_{seg} < -0.2$ eV).

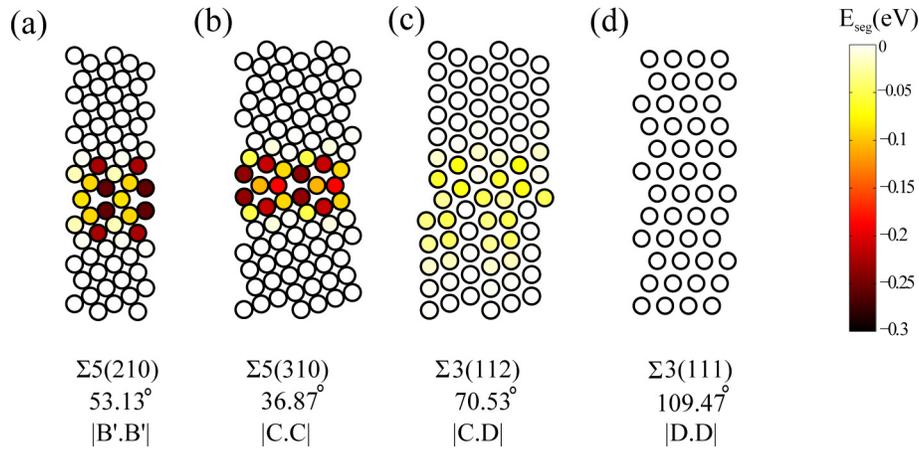

Fig. 7. Site-specific segregation energies for (a) Σ5(210), (b) Σ5(310), (c) Σ3(112) and (d) Σ3(111) GBs. The corresponding structural units and Voronoi volume are illustrated in Fig. 2.

The Ga segregation length scales (based on EAM MS calculations) were determined for all of the grain boundaries in a similar manner to the methodology adopted in the previous section. The length-scale can be correlated with the Ga wetting thickness typically observed in experimental studies [5,23]. There are a large number of sites within 10 atomic planes of the GB plane that exhibit significant segregation energies (Fig. 6). The overall characteristic Ga segregation length scale determined using FWHM was found to be about ~10, 8, and 12 layers for <100>, <110>, and <111> STGBs, respectively. This analysis provides a tool that could be used to perform GB engineering to produce materials that are less susceptible to LME.

Grain boundaries that are at the extremes of the Ga segregation spectrum are identified by plotting segregation energies onto a stereographic triangle representing the STGB planes, as in Fig. 1(b). The segregation energies are shown for $d/d_{\{hkl\}} = \{0,1,2,3,4,5\}$ on the STGB plane stereographic triangle in Fig. 8 (*cf.* Fig. 1(b)). These plots show that the segregation energies decay to zero with fewer atomic planes for the <100> and <110> boundaries as compared with the <111> boundaries and that there is a stronger tendency for segregation to $d/d_{\{hkl\}} = 1$ than to the boundary plane itself.

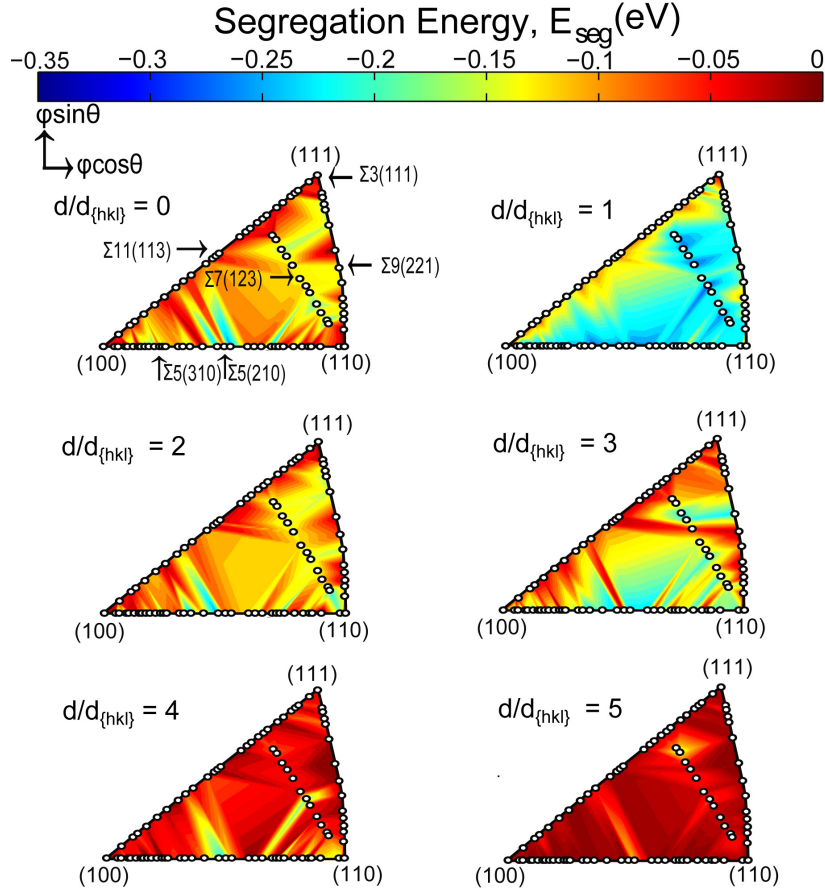

Fig. 8. The mean segregation energies at $d/d_{\{hkl\}}$ = {0,1,2,3,4,5} projected onto a stereographic triangle representing the STGB planes.

The boundaries with a weak ($E_{seg}^1 > -0.1$ eV) and a strong ($E_{seg}^1 \leq -0.26$ eV) Ga segregation tendency (based on the segregation energies at $d/d_{\{hkl\}}$=1) are tabulated in Tables III and IV, respectively. Table III consists of both a few small and multiple high angle GBs all with <110> tilt axes and which are vicinal to the coherent twin boundary, Σ3(111)<110>. The resistance of the coherent twin to liquid Ga penetration was previously reported [23]. Many of these Ga segregation resistant GBs contain a combination of the C and D structural units, which may be beneficial to LME. Since the facets of <110> asymmetric tilt GBs may contain these structural units [58], increasing the fraction of these beneficial GBs may aid in resisting LME. The GBs listed in Table IV shows that the boundaries with the most negative Ga segregation energies are all from either the <100> or <111> tilt axis families and are expected to show the least resistance to liquid Ga penetration. Glickman [79] suggested that crack velocity increases strongly with decreasing solid-liquid surface energies and an increasing GB energy, i.e., decreasing grain boundary cohesive energy. The segregation of Ga atoms to GBs can clearly influence grain boundary cohesive property of boundaries by reducing GB energies.

Further analyses of these simulations reveal a relationship between the vacancy binding and Ga segregation energies for the <100>, <110>, and <111> Al STGBs. The degree of correlation was analyzed with the help of linear correlation factor 'R', where $-1 \leq R \leq 1$ (R = 1 indicates a perfect positive correlation and R = -1 indicates a perfect anti-correlation). Table V lists a few STGBs with their corresponding correlation factor 'R'. All the calculations are statistically significant with $p<0.05$. There is a very strong correlation (R>0.7) for vacancy binding and Ga segregation for almost all boundaries. The exception is the Σ3(111) coherent twin boundary (where the magnitude of vacancy binding and segregation are very weak).

TABLE III. GBs which exhibit low segregation tendencies (i.e. $0.1\ eV < E^1_{seg}$) at $d/d_{\{hkl\}} = 1$. These GBs have a relatively low susceptibility to LME.

| GB Σ {(plane) <axis> | Misorientation (°) | X x Y x Z (Å) | Length Scale ($d/d_{\{hkl\}}$) | No of atoms | GB Energy γ (mJ/m²) | Seg. Energy $E^1_{seg}$ (eV) |
|---|---|---|---|---|---|---|
| Σ129 (1,1,16) <110> | 10.10 | 91.0 x 255.6 x 5.72 | 6 | 8104 | 310 | -0.06 |
| Σ33 (118) <110> | 20.05 | 46.5 x 262.1 x 5.72 | 6 | 4200 | 401 | -0.06 |
| Σ57 (558) <110> | 82.95 | 61.2 x 259.2 x 5.72 | 6 | 5464 | 402 | -0.07 |
| Σ17 (223) <110> | 86.63 | 23.6 x 267.3 x 5.72 | 6 | 2176 | 389 | -0.06 |
| Σ57 (445) <110> | 97.06 | 43.2 x 244.0 x 5.72 | 6 | 3640 | 334 | -0.05 |
| Σ43 (556) <110> | 99.37 | 53.1 x 298.6 x 5.72 | 4 | 5472 | 302 | -0.08 |
| Σ81 (778) <110> | 102.12 | 72.9 x 308.2 x 5.72 | 6 | 7752 | 254 | -0.06 |
| Σ131 (9,9,10) <110> | 103.69 | 92.7 x 259.3 x 5.72 | 4 | 8296 | 222 | -0.07 |
| Σ3 (111) <110> | 109.47 | 9.9 x 257.2 x 5.72 | 0 | 880 | 11 | 0.00 |
| Σ67 (776) <110> | 117.56 | 66.3 x 280.3 x 5.72 | 4 | 6400 | 248 | -0.06 |

TABLE IV. Selected GBs which exhibit a strong tendency for Ga segregation (i.e. $-0.26 \leq E_{seg}^1$) at $d/d_{\{hkl\}} = 1$. These GBs have a relatively high susceptibility to LME.

| GB Σ (plane)<axis> | Misorientation (°) | X x Y x Z (Å) | Length Scale ($d/d_{\{hkl\}}$) | # of atoms | GB Energy (mJ/m²) | Seg. Energy $E_{seg}^1$ (eV) |
|---|---|---|---|---|---|---|
| 169 (12,5,0) <100> | 45.24 | 52.6 x 315.8 x 4.05 | 6 | 4044 | 599 | -0.27 |
| 29 (730) <100> | 46.40 | 30.8 x 246.5 x 4.05 | 6 | 1848 | 595 | -0.26 |
| 43 (167) <111> | 15.18 | 65.0 x 301.18 x 7.01 | 6 | 8260 | 552 | -0.28 |
| 31 (156) <111> | 17.89 | 55.2 x 255.3 x 7.01 | 6 | 5948 | 540 | -0.26 |
| 291 (5,14,19) <111> | 29.51 | 169.2 x 301.18 x 7.01 | 6 | 27684 | 562 | -0.28 |
| 201 (5,11,16)<111> | 35.57 | 140.6 x 321.17 x 7.01 | 4 | 19074 | 548 | -0.30 |
| 7 (123) <111> | 24.43 | 26.2 x 241.97 x 7.01 | 6 | 2676 | 517 | -0.27 |
| 309 (7,13,20) <111> | 40.35 | 174.4 x 402.36 x 7.01 | 6 | 29592 | 512 | -0.30 |
| 93 (4,7,11) <111> | 42.10 | 95.7 x 326.36 x 7.01 | 4 | 13164 | 493 | -0.27 |
| 129 (5,8,13) <111> | 44.82 | 112.6 x 259.60 x 7.01 | 8 | 12330 | 457 | -0.27 |

TABLE V. GBs that exhibit a large correlation (R) between vacancy binding and segregation energies.

| GB Σ (plane) <axis> | Linear coefficient R |
|---|---|
| 5(210)<100> | 0.7917 |
| 5(310)<100> | 0.9137 |
| 13(320)<100> | 0.9249 |
| 13(510)<100> | 0.9150 |
| 17(410)<100> | 0.9455 |
| 3(112)<110> | 0.7864 |
| 3(111)<110> | -0.9934 |
| 11(332})<110> | 0.9331 |
| 9(221})<110> | 0.8500 |
| 11(113)<110> | 0.9114 |
| 21(145)<111> | 0.9597 |
| 57(178)<111> | 0.9567 |
| 7(123)<111> | 0.9047 |
| 13(134)<111> | 0.9307 |
| 129(5,8,13)<111> | 0.8717 |

## 6. Grain boundary cohesion

The accumulation of Ga near the GB plays a major role in GB fracture because it affects the GB cohesive energy. Ga segregation can also influence the cohesive energy by modifying the free energy of the free surfaces formed when a grain boundary separates. To examine this effect, we examined Ga segregation energies both near the GB and a free surface (FS). The results are shown in Fig. 9 for several GBs and their corresponding free surfaces, which shows that there is a greater tendency for Ga to segregate to the FS than to the GB (i.e., the segregation energy is more negative at the FS than GB). This result is similar to those seen in earlier studies for different metal systems such as Fe, Al, Cu with H and Fe with B, C, P and S [75,76]. For example, the Ga segregation energy in Al to a Σ5(210) STGB is approximately -0.07 eV and -0.36 eV at the FS, which is within 0.01 eV of values reported by Yamaguchi *et al*. [80]. These segregation energies suggest that (1) it is energetically favorable for Ga to segregate to the GB to lower its energy and (2) that is energetically more favorable to separate the grain boundary into two free surfaces in the presence of Ga segregation. Hence, these boundaries are susceptible to embrittlement. Furthermore, the embrittlement potency stems from both the ability of Ga to diffuse to the GB and its effect on the cohesive strength of the boundary (influenced by the difference between GB and FS segregation energies). Together, these criteria can be termed the segregation potency and embrittlement potency of Ga in particular GBs; both are required for LME. Interestingly, in the case of the $\Sigma 3(111)$ STGB, the segregation energy at the GB is low (~0 eV) while at the FS it is high (-0.21 eV). Even though this boundary supports GB embrittlement by Ga due to the embrittlement potency (difference between GB and FS segregation energies), the GB segregation potency indicates that there little driving for Ga to segregate to the boundary in the first place. Experimentally, this boundary has been associated with lower penetration rates [23], consistent with the results in Fig. 9, i.e., the energetics do not favor Ga segregation to this boundary.

With no applied stress, the susceptibility of GBs to intergranular failure can be analyzed on the basis of cohesive energies. The decrease in GB cohesive energies upon Ga segregations indicates that the system will embrittle with increasing Ga concentration at the boundary. In the case of a pure $\Sigma 5(210)$ boundary, the calculated GB energy and surface energy are 0.565 $J/m^2$ and 1.017 $J/m^2$, respectively, which yields a cohesive energy of 1.469 $J/m^2$ (Eq. 3). This is representative of the energy that must be supplied to separate the boundaries (per unit area). The calculated cohesive energy is within 1.4% of previously-simulated values from the literature (1.45 $J/m^2$) [80]. Gallium segregation reduces this value.

Gallium may be viewed as reducing GB cohesion, which aids in embrittling the GB. For cases where the FS segregation energies are significantly less negative than those for the GB, embrittlement is not energetically favored and the impurity enhances cohesion. Thus, Ga promotes decohesion for most of Al STGBs evaluated here (Fig. 9). In the systems where the tendency for GB segregation is very low (such as the coherent twin) and the tendency for FS segregation is significant, Ga has little effect on GB cohesion simply because it is energetically unfavorable for Ga to segregate to the GB in the first place.

Anderson *et al*. [81] experimentally explored the influence of Bi as an embrittler of selected Cu GBs ($\Sigma 11(113)$ STGB, $\Sigma 9(221)$ STGB, $\Sigma 5(310)$ STGB, ATGB) in bicrystals. Interestingly, they found that the ductility of the $\Sigma 11(113)$ STGB bicrystal with <100 ppm (0.003 at. %) Bi was not significantly affected after annealing at up to 773 K in a vacuum (to segregate Bi to boundaries), then a Bi vapor and then annealing in a liquid Bi bath. This behavior was unlike for the other bicrystals, which showed limited ductility after these annealing treatments. The Σ11(113)<110> STGB shows a high resistance to Bi segregation and to embrittlement, consistent with the present findings in Al that show the absence of GB segregation implies the absence of LME. Moreover, this lends support to the idea that boundaries with low segregation tendencies (see Table III) also resist liquid metal embrittlement.

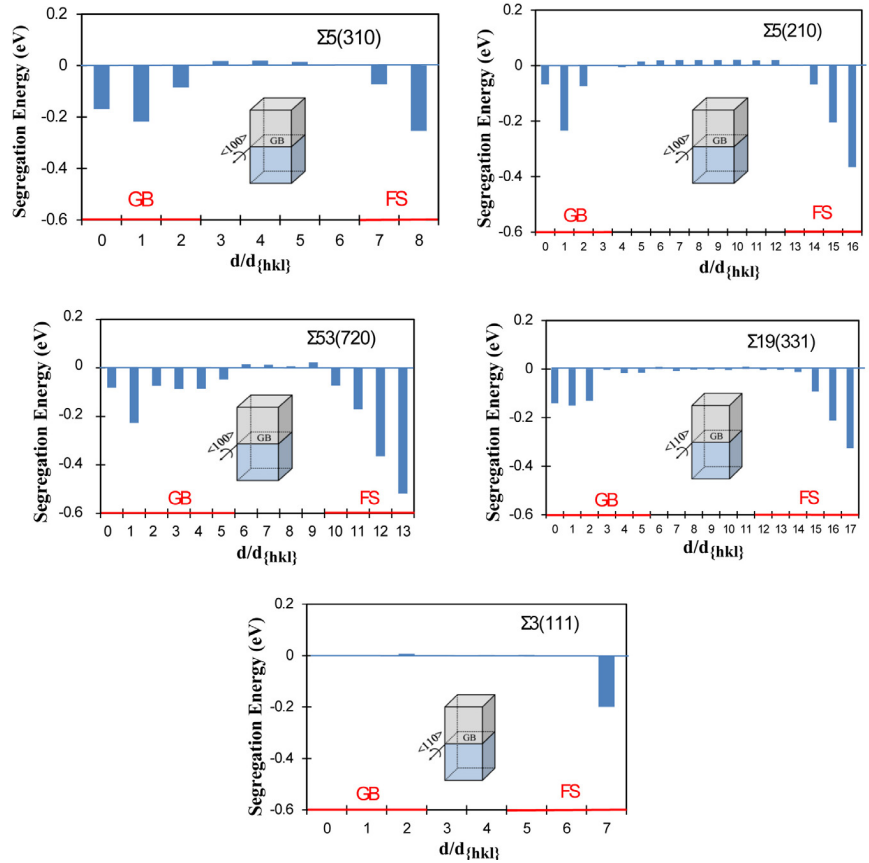

Fig. 9. Variation of segregation energy versus to distance from GB plane, $d/d_{\{hkl\}}$. The first entry in these plots is for the GB plane and the last is for the free surface for a few selected STGBs with <100> and <110> tilt axes.

## 7. Conclusion

In this work, we investigated the energetics of vacancy and Ga segregation to 125 different GBs in Al to provide insight into the mechanism of liquid-metal embrittlement. Simulations were performed using an EAM interatomic potential in a molecular statics simulation and via *ab initio* methods. We determined both the vacancy binding energies and Ga segregation energies in the vicinity of several symmetric tilt GBs in Al with <100>, <110>, and <111> tilt axes. We showed that GB structure and associated free volume play significant roles in Ga segregation and the subsequent embrittlement of Al. More specifically, this study shows that the local atomic structure of the GB and its structural units have a significant influence on the magnitude of vacancy formation energies. For example, the mean vacancy binding energy for <100>, <110>, and <111> STGBs (one atomic plane from the boundary) is -0.63 eV, -0.26 eV, and -0.60 eV, respectively (see Fig. 4). In general, the vacancy/grain boundary binding energies are large and negative, implying that vacancies are strongly bound to GBs in Al. The main exceptions were <110> boundaries that were either coherent twins or slightly vicinal with respect to the twin. Such boundaries do not provide good pathways for vacancy/substitutional diffusion. On average, the vacancy/GB binding energies approach zero 5 atomic planes from the GB plane for the <100> and <111> STGBs and 4 atomic planes for the <110> STGBs.

Ga segregation and its relationship to GB character provides a good starting place for grain boundary engineering to mitigate or reduce liquid-metal embrittlement in engineering alloys. Segregation tends to be strongest (most negative segregation energy) for sites close to the GB (Fig. 6). The mean Ga segregation energy for <100>, <110> and <111> STGBs (one atomic plane from the GB where the segregation is strongest) is -0.23 eV, -0.12 eV, and -0.24 eV, respectively, suggesting a strong correlation between the GB structural unit, its free volume, and segregation behavior, and could result in GB decohesion as reported by Ludwig and Bellet [16] and by Nam and Srolovitz [32]. A strong positive correlation between vacancy binding energy and Ga segregation energies was observed. The characteristic length scale association with Ga segregation was found to be ~10, 8, and 12 atomic planes for <100>, <110>, and <111> STGBs, respectively. This length scale may be related to the thickness of the wetting layers seen at GBs in the Ga/Al system, as observed experimentally in the Cu-Bi and Ni-Bi systems [1,17]. Several GBs were found for which the segregation energy was very small (low tendency for segregation). These boundaries could form the atomistic basis for grain boundary engineering to mitigate/control LME.

The perspectives presented here provide a physical basis for understanding the relationship between GB structure, Ga segregation and Ga transport kinetics as input to higher-scale modeling. For example, a phase-field model of impurity segregation would require both energetics and kinetic input to model the evolution of the solute distribution in a polycrystal. This will, in turn, provide better understanding of LME in alloys and ultimately a framework for engineering the GB character of polycrystals to modify embrittlement behavior in Al.

## Acknowledgements

The authors acknowledge useful insights and suggestions from Dr. W. Mullins and Dr. A.K. Vasudevan from the Office of Naval Research. This work was supported by the Office of Naval Research, contract No. N000141110793. MAT acknowledges the support of the U.S. Army Research Laboratory (ARL) administered by the Oak Ridge Institute for Science and Education through an interagency agreement between the U.S. Department of Energy and ARL.